\documentclass[
prd
,twocolumn
,superscriptaddress,noshowpacs,nofootinbib
,tightenlines
]{revtex4-1}
\usepackage{mathrsfs}
\usepackage{latexsym,bm}
\usepackage{graphicx}
\usepackage{indentfirst}
\usepackage{slashed}
\usepackage{amssymb}
\usepackage{amsmath}
\usepackage{bbm}
\usepackage{color}
\usepackage[dvipsnames]{xcolor}
\usepackage{epsfig}
\usepackage[titletoc]{appendix}
\usepackage{multirow}%
\usepackage{rotating}
\usepackage{epstopdf}
\usepackage{extarrows}
\usepackage{tabularx}
\usepackage{soul} 
\usepackage{xcolor}
\usepackage{lipsum}
\usepackage[colorlinks=true,allcolors=blue,hyperfootnotes=false]{hyperref}
\usepackage{bbold}

\usepackage[utf8]{inputenc}

\newcommand{\be}{\begin{equation}} \newcommand{\ee}{\end{equation}}
\newcommand{\ba}{\begin{array}{c}} \newcommand{\ea}{\end{array}}
\newcommand{\bea}{\begin{eqnarray}} \newcommand{\eea}{\end{eqnarray}}

\allowdisplaybreaks

\begin{document}

\title{\LARGE Neutral-current weak pion production off the nucleon\\
in covariant chiral perturbation theory
}
\author{De-Liang Yao}
\email{yaodeliang@hnu.edu.cn}
\affiliation{School of Physics and Electronics, Hunan University, Changsha 410082, China}
\affiliation{Departamento de F\'{\i}sica Te\'orica and Instituto de F\'{\i}sica Corpuscular (IFIC), Centro Mixto CSIC-UV, Institutos de Investigaci\'on de Paterna,
E-46071, Valencia, Spain}
\author{Luis Alvarez-Ruso}
\email{Luis.Alvarez@ific.uv.es}
\affiliation{Departamento de F\'{\i}sica Te\'orica and Instituto de F\'{\i}sica Corpuscular (IFIC), Centro Mixto CSIC-UV, Institutos de Investigaci\'on de Paterna,
E-46071, Valencia, Spain}
\author{M. J.~Vicente Vacas}
\email{Manuel.J.Vicente@ific.uv.es}
\affiliation{Departamento de F\'{\i}sica Te\'orica and Instituto de F\'{\i}sica Corpuscular (IFIC), Centro Mixto CSIC-UV, Institutos de Investigaci\'on de Paterna,
E-46071, Valencia, Spain}

\date{\today}
\begin{abstract}
Neutral current single pion production induced by neutrinos and antineutrinos on nucleon targets has been investigated in manifestly relativistic baryon chiral perturbation theory with explicit $\Delta(1232)$ degrees of freedom up to $\mathcal{O}(p^3)$. At low energies, where chiral perturbation theory is applicable, the total cross sections for the different reaction channels exhibit a sizable non-resonant contribution, which is not present in event generators of broad use in neutrino oscillation and cross section experiments such as GENIE and NuWro. 
\end{abstract}
\maketitle

\section{Introduction}
In recent years, more precise measurements of neutrino cross sections have been achieved, and more experiments are still ongoing or planned to enrich the wealth of data for a better determination of the mechanisms of neutrino interactions with matter (see for instance Ref.~\cite{Mahn:2018mai} and references therein). A better understanding of neutrino interactions with nucleons and nuclei is of particular importance to achieve the precision goals of modern neutrino-oscillation experiments~\cite{Benhar:2015wva,Alvarez-Ruso:2017oui}. 

Neutrino- and antineutrino-induced single pion production are among the dominant contributions to the inclusive (anti)neutrino-nucleus cross section in the relevant energy regime of several experiments. In particular, neutral current (NC) $\pi^0$ production can mimic the electron-like signal in $\nu_\mu \to \nu_e$ ($\bar{\nu}_\mu \to \bar{\nu}_e$)  measurements. Currently, uncertainties at the $20-30\%$ level have to be considered for single pion production in the analyses of oscillation experiments, owing to the conflict between data sets and models~\cite{Katori:2016yel}. Single pion production off the nucleon, as a fundamental ingredient of the corresponding process on nuclei should be investigated.

Weak single pion production off the nucleon has been extensively studied at intermediate energies with various phenomenological models. Models accounting for the $\Delta(1232)$ resonance and higher nucleon excitations have been developed~\cite{Adler:1968tw,Bijtebier:1970ku,LlewellynSmith:1971uhs,Alevizos:1977xf,Fogli:1979cz,Rein:1980wg,Hernandez:2007qq,Leitner:2008ue,Barbero:2008zza,Sato:2009de,Serot:2012rd,Graczyk:2014dpa,Nakamura:2015rta,Alam:2015gaa,Gonzalez-Jimenez:2016qqq,Kabirnezhad:2017jmf}, with some studies specifically addressing the NC case~\cite{Fogli:1979qj,Leitner:2006sp}. In addition to nucleon excitation, most of these models incorporate non-resonant amplitudes with form factors in the vertices. 

Recently, a well founded low-energy model for charged-current (CC) weak single pion production off the nucleon has been developed in Ref.~\cite{Yao:2018pzc} on the basis of covariant baryon chiral perturbation theory (ChPT)~\cite{Weinberg:1978kz,Gasser:1983yg,Gasser:1984gg,Bernard:1995dp,Scherer:2012xha}. The study has been systematically performed up to chiral order $\mathcal{O}(p^3)$, where loop diagrams are present. The extended-on-mass-shell (EOMS) scheme~\cite{Fuchs:2003qc} was adopted to remedy the power counting breaking of ChPT in the presence of baryons~\cite{Gasser:1987rb}. The EOMS scheme has been extensively used in various studies of baryonic phenomenology, see e.g. Refs.~\cite{Geng:2008mf,Alarcon:2012kn,Chen:2012nx,Hilt:2013fda,Yao:2016vbz,Blin:2016itn,Siemens:2017opr,Yao:2017fym}, even beyond the low-energy region~\cite{Epelbaum:2015vea}.\footnote{The heavy-baryon approach has also proved to be adequate for pion production off the nucleon induced by vector, see, e.g., Refs.~\cite{FernandezRamirez:2012nw,Hilt:2013uf}, and  also axial~\cite{Bernard:1993xh,Yao:2018pzc} currents, at least very close to threshold.  } Furthermore, the $\Delta$ resonance was explicitly included according to the $\delta$-counting rule from Ref.~\cite{Pascalutsa:2002pi}. 

In this letter, we extend the study to the neutral-current reactions. For this purpose, new isoscalar amplitudes, not present in the CC case and involving three distinctive low-energy constants (LECs), need to be computed. Following a general description of the electroweak amplitude for NC weak pion production on the nucleon, the external currents and the 
relevant terms of the chiral Lagrangian are introduced. The calculation of the hadronic matrix elements is then briefly described. Numerical results for integrated cross sections at low energies are presented in comparison to the output of neutrino event generators and phenomenological models. 

\section{Neutral-current single pion production\label{sec:NCspp}}

Neutrino-induced single-pion production off the nucleon is represented by 
\bea\label{eq:piprod}
\nu(k_1)+N(p_1)\to \nu(k_2)+N^\prime(p_2)+\pi^b(q)\ ,
\eea
with the four momenta of all particles indicated in parenthesis; $b$ denotes the pion isospin index.  In the standard model, this process is mediated by the vector $Z$-boson, propagating with mass $M_Z$ and transferred momentum squared $t_1=(k_1-k_2)^2$. Under the one-boson exchange approximation and in the limit of $t_1\ll M_Z^2$, the Lorentz-invariant amplitude is cast as
\bea\label{eq:T}
\mathcal{T}=\frac{G_F}{\sqrt{2}}\big[\bar{u}(k_2)\gamma_\alpha(1-\gamma_5)u(k_1)\big]\langle \pi^b N^\prime|\mathcal{J}^\alpha(0)|N\rangle ,
\eea
where $G_F=1.166 \times 10^{-5}$~GeV$^{-2}$ is the Fermi constant. The factor in square brackets is the leptonic current, while the one after is the neutral current hadronic matrix element. Its isospin structure  has the form
\bea
\langle \pi^b N^\prime|\mathcal{J}_\alpha(0)|N\rangle=\chi_f^\dagger\big[\delta^{b3}{H}_\alpha^++i\epsilon^{b3c}\tau^c{H}_\alpha^-+\tau^b{H}_\alpha^0\big]\chi_i\ .\nonumber
\eea
Here $\tau^b$ denotes a Pauli matrix; $\chi_{i,f}$ are isospinors of the initial and final nucleons.  ${H}_\alpha^{\pm}$ stand for the isospin even/odd amplitudes, responsible for the isovector part of the NC. Likewise, $H_\alpha^0$ is the isoscalar amplitude. These isospin amplitudes  are often further recast in terms of Lorentz vector  ($V_\alpha$) and axial-vector ($A_\alpha$) amplitudes, as 
\bea
{H}_\alpha^{\pm}&=&\big(1-2\sin^2\theta_W\big)V_\alpha^\pm - A_\alpha^\pm\ ,\nonumber\\
{H}_\alpha^{0}&=&\big(-2\sin^2\theta_W\big)V_\alpha^{0}\ ,
\eea
with $\sin\theta_W\simeq0.464$, being the sine of the weak angle $\theta_W$. 
Isospin symmetry implies that $V^\pm_\alpha$ and $A^\pm_\alpha$ are also present in the corresponding CC reactions\footnote{Namely, in the CC case the amplitudes $H^{\pm (\mathrm{CC})}_\alpha$ defined by Eq.~(8) of Ref.~\cite{Yao:2018pzc} are given by $H^{\pm (\mathrm{CC})}_\alpha =\sqrt{2} (V_\alpha^\pm - A_\alpha^\pm)$.}. 
Hadronic amplitudes  for all the physical processes can be obtained from the isospin ones $H_\alpha^{\pm,0}$ via
\bea
H_\alpha(\nu p\to\nu p\pi^0)&=&H_\alpha^++H_\alpha^0\ ,\nonumber\\
H_\alpha(\nu n\to\nu n\pi^0)&=&H_\alpha^+-H_\alpha^0\ ,\nonumber\\
H_\alpha(\nu n\to\nu p\pi^-)&=&\sqrt{2}(H_\alpha^0 - H_\alpha^-)\ ,\nonumber\\
H_\alpha(\nu p\to\nu n\pi^+)&=&\sqrt{2}(H_\alpha^0 + H_\alpha^-)\ .\label{eq:phys}
\eea

For the antineutrino-induced reactions, one just needs to substitute $\nu\to \bar{\nu}$ in Eqs~\eqref{eq:piprod} and~\eqref{eq:phys} and, correspondingly, $u(k_1)\to v(k_2)$, $\bar{u}(k_2) \to \bar{v}(k_1)$ in Eq.~\eqref{eq:T}.  

\section{Hadronic amplitudes from ChPT}
We proceed by deriving the hadronic amplitudes in ChPT, aiming at a systematic description of  NC weak pion production at low energies. In Ref.~\cite{Yao:2018pzc}, the CC single pion production has been studied in covariant ChPT with explicit $\Delta$ resonances up to $\mathcal{O}(p^3)$. The same framework therein can be extended to the NC case; the chiral Lagrangian should be now coupled to the following external fields:
\bea 
l_\mu &=& \left(\frac{g_W}{2 \cos{\theta_W}}\right) (-2 \cos^2{\theta_W}) Z_\mu \frac{\tau^3}{2} \,, \label{eq.lc}\\ 
r_\mu &=& \left(\frac{g_W}{2 \cos{\theta_W}}\right)  (2 \sin^2{\theta_W}) Z_\mu \frac{\tau^3}{2} \,, \label{eq.rc}\\ 
v^{s}_\mu &=& \left(\frac{g_W}{2 \cos{\theta_W}}\right) \sin^2{\theta_W} Z_\mu\,\mathbb{1}_{2\times 2} \ ,
\eea
with $g_W$ the weak coupling constant. 
The common factor $g_W/(2 \cos{\theta_W})$ is factorized from $H_\alpha^{\pm, 0}$ and, combined with an identical factor from the leptonic vertex, written in terms of $G_F$ in Eq.~(\ref{eq:T}). Left and right fields contribute to $H_\alpha^{\pm}$ which, in practice, can be directly taken from the CC calculation of Ref.~\cite{Yao:2018pzc} thanks to isospin symmetry, as explained in the previous section. The additional terms in the chiral Lagrangian required for the NC calculation up to $\mathcal{O}(p^3)$ are
{
\bea
\mathcal{L}_{\pi N}^{(1)} &\supset&  \bar{\Psi}_N \gamma^\mu v^s_\mu \Psi_N \,, \\
\mathcal{L}_{\pi N}^{(2)} &\supset&  \bar{\Psi}_N \frac{1}{4m} \left(c_6+2c_7\right)\,v^s_{\mu\nu} \sigma^{\mu\nu} \Psi_N \,, 
\label{eq:c67}
\\
\mathcal{L}_{\pi N}^{(3)} &\supset&  \bar{\Psi}_N \bigg[\frac{2d_7}{m}i[D^\mu,v^s_{\mu\nu}]D^\nu\nonumber\\
&&\hspace{0.6cm}+\frac{2d_9}{m}i\epsilon^{\mu\nu\alpha\beta}  v^s_{\mu\nu} u_\alpha D_\beta+{\rm h.c.}\bigg] \Psi_N\ ,\label{eq:Lag_OLD}
\eea
where $v^s_{\mu\nu} = \partial_\mu v^s_\nu - \partial_\nu v^s_\mu$; $\Psi_N=(p,n)^T$ represents the nucleon doublet and $m$ is the nucleon mass in the chiral limit. Here the parameters $c_7$, $d_7$ and $d_9$ are additional LECs which do not appear in CC production reactions. 
The above Lagrangian terms are extracted from Ref.~\cite{Fettes:2000gb} by singling out the isoscalar current. For this purpose, the current tensor $F_{\mu\nu}^+$ and the covariant derivative $D_\mu$ in that paper have been split as follows: 
\bea
F_{\mu\nu}^+&=&h^+_{\mu\nu}\tau^3+2v_{\mu\nu}^{s}\ ,\\
D_\mu\Psi_N&=&(\partial_\mu+\Gamma_\mu-iv_\mu^s)\Psi_N\ ,
\eea
where the isovector piece $h_{\mu\nu}^+\tau^3$ is constructed from $l_\mu$ and $r_\mu$ in Eqs.~\eqref{eq.lc} and~\eqref{eq.rc}. We refer the readers to Ref.~\cite{Fettes:2000gb} for the explicit expressions of the chiral blocks.}

The calculation of $H_\alpha^0$ can be readily carried out following the procedure demonstrated in detail in Ref.~\cite{Yao:2018pzc} and using the same topologies of Feynman diagrams. We apply the $\delta$-counting rule~\cite{Pascalutsa:2002pi} due to the inclusion of explicit $\Delta$ resonances and the low energies considered.  Like in the CC case, the obtained analytical expressions are too lengthy to be displayed explicitly here. They are available from the authors upon request. We have  checked that all the ultraviolet (UV) divergences and power counting breaking (PCB) terms in loops can be properly canceled by the LECs in the chiral Lagrangians. Interestingly, it turns out that the sum of the isoscalar loop amplitudes does not suffer from UV divergences, indicating that the $\beta$ functions for the new LECs $c_7$, $d_7$ and $d_9$ are equal to zero. Namely,
\bea
X=X^r\ ,\qquad X\in\{c_7,d_7,d_9\}\ ,
\eea
where $X^r$ are the UV-renormalized counterparts for the parameters $X$. In order to remove the PCB terms one has to further make the finite shift~\footnote{For $c_6$, also present in the CC case~\cite{Yao:2018pzc}, the shift is 
$c_6^r=\tilde{c}_6-\frac{5g^2m^2}{16\pi^2F^2}$.
}
\be
c_7^r=\tilde{c}_7+\frac{g^2m^2}{4\pi^2F^2}\ ,
\ee
 where $\tilde{X}$ are the EOMS-renormalized LECs; $g$ and $F$ denote the axial coupling and pion decay constant in the chiral limit, respectively. Obviously, at $\mathcal{O}(p^3)$,  $d_7$ and $d_9$ are untouched by the cancellation of the PCB terms and hence $X^r=\tilde{X}$ for $X\in\{d_7,d_9\}$. All the other LECs are renormalized as described in section~III.D of Ref.~\cite{Yao:2018pzc}.

\section{Total cross section and numerical results}
In the center-of-mass (CM) frame of the initial (anti)neutrino-nucleon pair, the total cross section takes the form
\bea\label{eq:totCS}
\sigma(s)&=&\frac{1}{2 (4\pi)^4\sqrt{s}\,|\mathbf{k}_1|}\int_{\omega_\nu^-}^{\omega_\nu^+}{\rm d}\omega_\nu\int_{\omega_\pi^-}^{\omega_\pi^+}{\rm d}\omega_\pi\int_{-1}^{+1}{\rm d}x_1\nonumber\\
&&\times\int_0^{2\pi} {\rm d}\phi_{12}\,|{\cal T}|^2\,,
\eea
where $s\equiv (k_1+p_1)^2$, $x_1\equiv \cos\theta_1$, with $\theta_1$ the angle between $\vec{k}_1$ and $\vec{k}_2$; $\phi_{12}$ is the angle between the $\vec{k}_1\wedge \vec{k}_2$ plane and the $\vec{k}_2\wedge \vec{q}$ plane (see Figure~1 of Ref.~\cite{Yao:2018pzc} for clarification). Furthermore, $\omega_\nu$ and $\omega_\pi$ are the energies of the outgoing (anti)neutrino and pion, respectively.  Their kinematic limits are given by
\bea
&&\omega_\nu^-=0\ ,\quad \omega_\nu^+=\frac{(\sqrt{s}-M_\pi)^2-m_N^2}{2(\sqrt{s}-M_\pi)}\ ,\\
&&\omega_\pi^{\pm}=\frac{(\sqrt{s}-\omega_\nu) \Delta_s^+ \pm\,\omega_\nu\sqrt{(\Delta_s^-)^2-4M_\pi^2m_N^2}}{2(s-2\omega_\nu\sqrt{s})}\,,
\eea
with $\Delta_s^\pm=s-2\omega_\nu\sqrt{s} \pm M_\pi^2-m_N^2$, where $M_\pi$ and $m_N$ are the physical masses of the pion and the nucleon, respectively. Finally, the unpolarized Lorentz-invariant $\mathcal{T}$-matrix squared can be written as
\bea
|\mathcal{T}|^2=\frac{G_F^2}{2}L^{\alpha\beta} H_{\alpha\beta}\ ,
\eea
with the leptonic tensor ($\epsilon_{0123}=-1$),
\bea
L_{\alpha\beta}
=8[k_{1,\alpha} k_{2,\beta}+k_{1,\beta} k_{2,\alpha}-g_{\alpha\beta}k_1\cdot k_2\pm i\epsilon_{\alpha\beta\rho\sigma}k_{1}^{\rho}k_{2}^{\sigma}]\ .
\nonumber\eea
The plus (minus) sign corresponds to the neutrino- (antineutrino-) induced reactions, respectively. In terms of the hadronic amplitudes $H_\alpha$ introduced in Sec.~\ref{sec:NCspp}, the hadronic tensor reads
\bea
H_{\alpha\beta}&=&\frac{1}{2}{\rm Tr}\big[(\slashed{p}_1+m_N)\tilde{H}_\alpha(\slashed{p}_2+m_N)H_\beta\big]\ ,
\eea
where $\tilde{H}_\alpha=\gamma_0H^\dagger_\alpha\gamma_0$.

In our numerical computation, the values of those parameters common to both CC and NC pion production are assigned as in Ref.~\cite{Yao:2018pzc}. The extra LECs present in the NC case are fixed as follows:\footnote{The $c_7$ used here is related to the $c_6^{\rm BBS}$ and $c_7^{\rm BBS}$ in Ref.~\cite{Bauer:2012pv} by $c_7=m_N(c_7^{\rm BBS}-2\,c_6^{\rm BBS})$, where the superscripts 'BBS' have been added for the sake of clarity.} $c_7\simeq (-2.68\pm0.08)$, $d_7=-0.49$~GeV$^{-2}$ and $d_{9}=(0\pm 1)$~GeV$^{-2}$. More specifically, $c_7$ is determined using the empirical  values of the anomalous magnetic moments of the proton ($\kappa_p$) and the neutron ($\kappa_n$)~\cite{Patrignani:2016xqp,Bauer:2012pv}, while the error is a rough estimate of higher-order contributions based on the results of Ref.~\cite{Bauer:2012pv}. The value of $d_7$ is extracted from the proton and neutron electromagnetic radii in Ref.~\cite{Fuchs:2003ir}.\footnote{Likewise, since $d_6$ is also present in the NC case, we employ $d_6=-0.70$~GeV$^{-2}$ from Ref.~\cite{Fuchs:2003ir} although it was previously let undetermined and set to a natural size for CC pion production in Ref.~\cite{Yao:2018pzc}. } For the LEC $d_9$ there is no determination using the present framework. Therefore, it has been set to a natural value, which is a reasonable estimate only when the $\Delta$ resonance is explicitly taken into account, as in this study.

\begin{figure*}[ht!]
\begin{center}
\epsfig{file=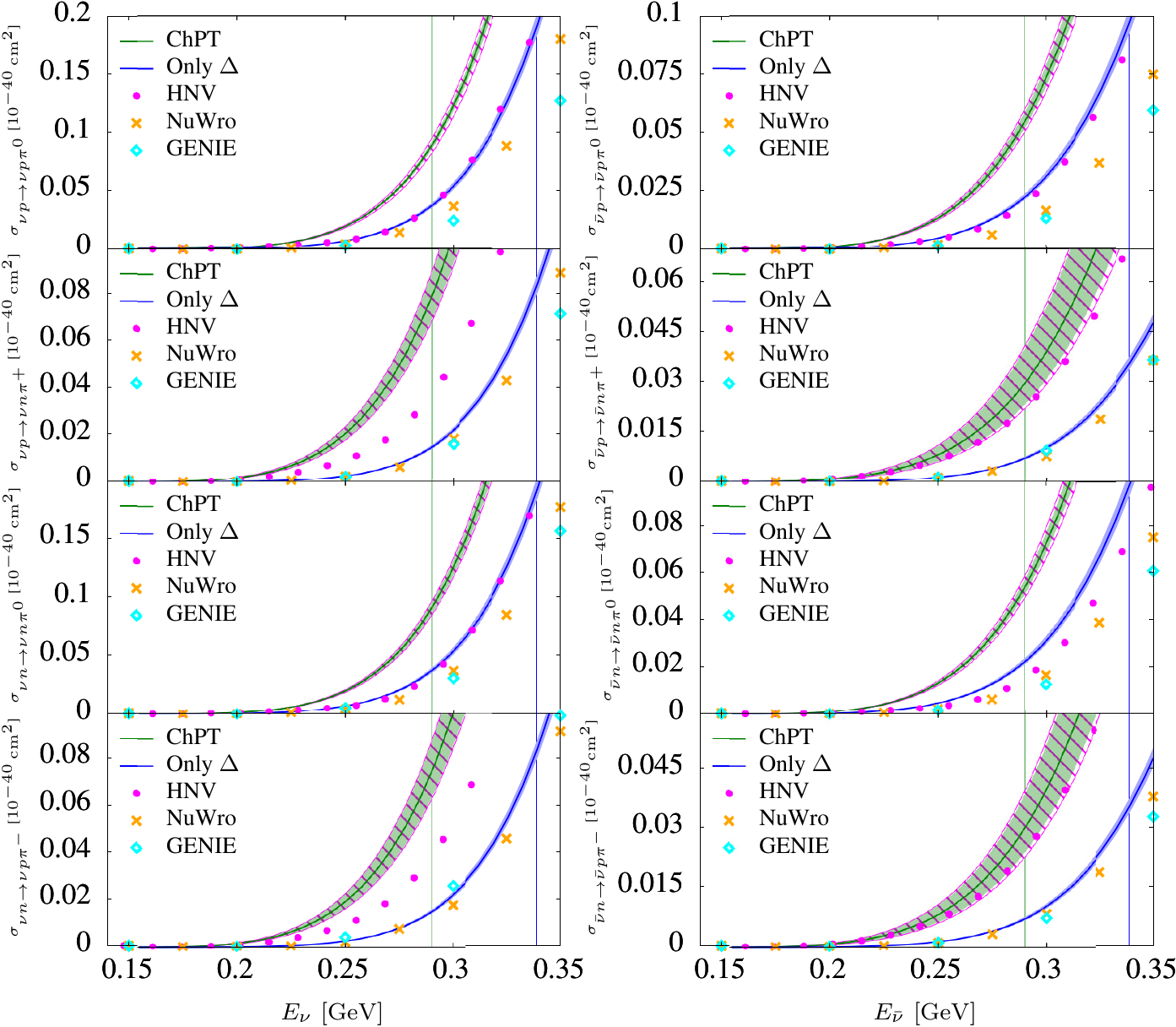,width=0.95\textwidth}
\caption{Cross sections for neutral-current weak pion production off the nucleon as a function of the Laboratory neutrino energy. The solid line represents the full $\mathcal{O}(p^3)$ ChPT prediction, while the dashed line stands for the ChPT result obtained with $\Delta$-exchange contributions alone. Our ChPT calculation is expected to be reliable up to the energy marked by 
the vertical green line. The vertical blue line at a higher energy indicates the energy at which the $\Delta$ pole starts to be reached. The magenta-hatched bands in the full results denote the total uncertainties from both higher-order truncation and LECs, while the green and blue bands account for 1$\sigma$ uncertainties propagated from the errors of the LECs. For comparison, results produced by the HNV model~\cite{Hernandez:2007qq}, NuWro~\cite{Juszczak:2005zs} and GENIE~\cite{Andreopoulos:2009rq} Monte Carlo generators are also shown by dots, crosses and hollow diamonds, respectively. }
\label{fig:cs.nc}
\end{center}
\end{figure*}
In Fig.~\ref{fig:cs.nc}, total cross sections for all the physical NC single pion production channels are displayed as a function of $E_\nu$, the energy of the incoming (anti)neutrino in the laboratory frame. As in the CC case~\cite{Yao:2018pzc}, our chiral predictions for the NC cross sections are expected to be reliable up to the neutrino energy $E_{\nu,{\rm ChPT}} \sim E_{\nu,{\rm th}}+M_\pi\simeq 283$~MeV, where the threshold energy
$
E_{\nu,{\rm th}}=M_\pi+{M_\pi^2}/{(2m_N)}\simeq 149.3~{\rm MeV}
$.
This energy interval $(E_{\nu,{\rm th}},E_{\nu,{\rm th}}+M_\pi)$ is relatively far away from $E_\nu^\Delta\simeq 338.8~{\rm MeV}$, at which the final $\pi N$ system starts reaching the $\Delta$ pole, so that the adopted $\delta$-counting rule is appropriate. Although of higher order, the $\Delta$-width has been incorporated in the $\Delta$ propagator as in Eq.~(A4) of Ref.~\cite{Yao:2018pzc}. This allows to extend the results smoothly to higher energies with minor impact on the cross sections in the energy region we are concerned with. The green and blue bands in the curves of Fig. 1 denote the statistical uncertainties propagated from the errors of the involved LECs.
The uncertainty due to the missing higher-order terms in the chiral expansion has been estimated using the method proposed in Refs.~\cite{Epelbaum:2014efa,Binder:2015mbz}.  
This error has been added in quadrature to the one propagated from the uncertainties in the LECs. Its effects are small. The full errors (statistical+higher order) for the total cross sections are represented by the magenta-hatched bands in Fig.~\ref{fig:cs.nc}.

For comparison, we also show the total cross sections produced by the NuWro~\cite{Juszczak:2005zs} and GENIE~\cite{Andreopoulos:2009rq} neutrino Monte Carlo generators. It can be observed from Fig.~\ref{fig:cs.nc} that both the NuWro and GENIE results agree with the ChPT ones with only the $\Delta$ contribution. However, they  underestimate the full ChPT predictions. The later reveal a sizable non-resonant contribution close to threshold, which is not accounted by the event generators.

Furthermore, we confront our results with a theoretical model proposed in Ref.~\cite{Hernandez:2007qq} and improved in Refs.~\cite{Hernandez:2013jka,Alvarez-Ruso:2015eva,Hernandez:2016yfb}, as described in Ref.~\cite{Sobczyk:2018ghy}. In that model, denoted as HNV in Fig.~\ref{fig:cs.nc}, on top of some resonances, the chiral non-resonant terms at lowest order are considered, which leads to  larger cross sections than in the NuWro and GENIE codes. On the other hand, our systematic ChPT calculation up to $\mathcal{O}(p^3)$ contains additional higher-order terms of $\mathcal{O}(p^2)$ and $\mathcal{O}(p^3)$, as well as loops. The HNV model only partially and phenomenologically accounts for higher-order contribution through empirical form factors and Watson phases. The $\mathcal{O}(p^3)$ ChPT calculation produces considerably larger cross sections  with respect to the HNV model in all reaction channels, with the  closest accord found for $\bar{\nu}n\to \bar{\nu}p\pi^-$. Similar conclusions are expected in the comparison to  the dynamical coupled-channel model of Refs.~\cite{Sato:2003rq,Matsuyama:2006rp,Nakamura:2015rta} since their results are in good agreement with the HNV model at low energies, as pointed out in Ref.~\cite{Sobczyk:2018ghy}.

\section{Summary}

We have studied neutral-current weak pion production off the nucleon in covariant ChPT with explicit $\Delta(1232)$ up to $\mathcal{O}(p^3)$ following the $\delta$-counting rule. The amplitudes have been renormalized using the EOMS scheme. Three new LECs not present in the CC case are now required.        By setting the LECs either to values determined elsewhere within the same scheme or, otherwise, to natural values, we have predicted the cross sections for all the eight physical processes induced by neutrinos or antineutrinos. In the incoming-neutrino energy interval in which the calculation is expected to be valid, our predictions for the $\Delta$ excitation mechanism conform well with the corresponding output from the widely used NuWro and GENIE neutrino event generators. However, for all the channels the full ChPT results are significantly larger due to the systematic and model-independent inclusion of non-resonant terms. This observation implies that a precise description of low-energy weak pion production requires a realistic account of non-resonant amplitudes.

\acknowledgments
We would like to thank J.~E.~Sobczyk for providing us the results for the HNV model. This research has been supported by the Spanish Ministerio de Econom\'ia y Competitividad (MINECO) and the European Regional Development Fund (ERDF), under contracts 
FIS2017-84038-C2-1-P, FIS2017-84038-C2-2-P, SEV-2014-0398.

\appendix

\bibliography{weak}

\begin{thebibliography}{57}%
\makeatletter
\providecommand \@ifxundefined [1]{%
 \@ifx{#1\undefined}
}%
\providecommand \@ifnum [1]{%
 \ifnum #1\expandafter \@firstoftwo
 \else \expandafter \@secondoftwo
 \fi
}%
\providecommand \@ifx [1]{%
 \ifx #1\expandafter \@firstoftwo
 \else \expandafter \@secondoftwo
 \fi
}%
\providecommand \natexlab [1]{#1}%
\providecommand \enquote  [1]{``#1''}%
\providecommand \bibnamefont  [1]{#1}%
\providecommand \bibfnamefont [1]{#1}%
\providecommand \citenamefont [1]{#1}%
\providecommand \href@noop [0]{\@secondoftwo}%
\providecommand \href [0]{\begingroup \@sanitize@url \@href}%
\providecommand \@href[1]{\@@startlink{#1}\@@href}%
\providecommand \@@href[1]{\endgroup#1\@@endlink}%
\providecommand \@sanitize@url [0]{\catcode `\\12\catcode `\$12\catcode
  `\&12\catcode `\#12\catcode `\^12\catcode `\_12\catcode `\%12\relax}%
\providecommand \@@startlink[1]{}%
\providecommand \@@endlink[0]{}%
\providecommand \url  [0]{\begingroup\@sanitize@url \@url }%
\providecommand \@url [1]{\endgroup\@href {#1}{\urlprefix }}%
\providecommand \urlprefix  [0]{URL }%
\providecommand \Eprint [0]{\href }%
\providecommand \doibase [0]{http://dx.doi.org/}%
\providecommand \selectlanguage [0]{\@gobble}%
\providecommand \bibinfo  [0]{\@secondoftwo}%
\providecommand \bibfield  [0]{\@secondoftwo}%
\providecommand \translation [1]{[#1]}%
\providecommand \BibitemOpen [0]{}%
\providecommand \bibitemStop [0]{}%
\providecommand \bibitemNoStop [0]{.\EOS\space}%
\providecommand \EOS [0]{\spacefactor3000\relax}%
\providecommand \BibitemShut  [1]{\csname bibitem#1\endcsname}%
\let\auto@bib@innerbib\@empty
\bibitem [{\citenamefont {Mahn}\ \emph {et~al.}(2018)\citenamefont {Mahn},
  \citenamefont {Marshall},\ and\ \citenamefont {Wilkinson}}]{Mahn:2018mai}%
  \BibitemOpen
  \bibfield  {author} {\bibinfo {author} {\bibfnamefont {K.}~\bibnamefont
  {Mahn}}, \bibinfo {author} {\bibfnamefont {C.}~\bibnamefont {Marshall}}, \
  and\ \bibinfo {author} {\bibfnamefont {C.}~\bibnamefont {Wilkinson}},\ }\href
  {\doibase 10.1146/annurev-nucl-101917-020930} {\bibfield  {journal} {\bibinfo
   {journal} {Ann. Rev. Nucl. Part. Sci.}\ }\textbf {\bibinfo {volume} {68}},\
  \bibinfo {pages} {105} (\bibinfo {year} {2018})},\ \Eprint
  {http://arxiv.org/abs/1803.08848} {arXiv:1803.08848 [hep-ex]} \BibitemShut
  {NoStop}%
\bibitem [{\citenamefont {Benhar}\ \emph {et~al.}(2017)\citenamefont {Benhar},
  \citenamefont {Huber}, \citenamefont {Mariani},\ and\ \citenamefont
  {Meloni}}]{Benhar:2015wva}%
  \BibitemOpen
  \bibfield  {author} {\bibinfo {author} {\bibfnamefont {O.}~\bibnamefont
  {Benhar}}, \bibinfo {author} {\bibfnamefont {P.}~\bibnamefont {Huber}},
  \bibinfo {author} {\bibfnamefont {C.}~\bibnamefont {Mariani}}, \ and\
  \bibinfo {author} {\bibfnamefont {D.}~\bibnamefont {Meloni}},\ }\href
  {\doibase 10.1016/j.physrep.2017.07.004} {\bibfield  {journal} {\bibinfo
  {journal} {Phys. Rept.}\ }\textbf {\bibinfo {volume} {700}},\ \bibinfo
  {pages} {1} (\bibinfo {year} {2017})},\ \Eprint
  {http://arxiv.org/abs/1501.06448} {arXiv:1501.06448 [nucl-th]} \BibitemShut
  {NoStop}%
\bibitem [{\citenamefont {Alvarez-Ruso}\ \emph {et~al.}(2018)\citenamefont
  {Alvarez-Ruso} \emph {et~al.}}]{Alvarez-Ruso:2017oui}%
  \BibitemOpen
  \bibfield  {author} {\bibinfo {author} {\bibfnamefont {L.}~\bibnamefont
  {Alvarez-Ruso}} \emph {et~al.},\ }\href {\doibase 10.1016/j.ppnp.2018.01.006}
  {\bibfield  {journal} {\bibinfo  {journal} {Prog. Part. Nucl. Phys.}\
  }\textbf {\bibinfo {volume} {100}},\ \bibinfo {pages} {1} (\bibinfo {year}
  {2018})},\ \Eprint {http://arxiv.org/abs/1706.03621} {arXiv:1706.03621
  [hep-ph]} \BibitemShut {NoStop}%
\bibitem [{\citenamefont {Katori}\ and\ \citenamefont
  {Martini}(2018)}]{Katori:2016yel}%
  \BibitemOpen
  \bibfield  {author} {\bibinfo {author} {\bibfnamefont {T.}~\bibnamefont
  {Katori}}\ and\ \bibinfo {author} {\bibfnamefont {M.}~\bibnamefont
  {Martini}},\ }\href {\doibase 10.1088/1361-6471/aa8bf7} {\bibfield  {journal}
  {\bibinfo  {journal} {J. Phys.}\ }\textbf {\bibinfo {volume} {G45}},\
  \bibinfo {pages} {013001} (\bibinfo {year} {2018})},\ \Eprint
  {http://arxiv.org/abs/1611.07770} {arXiv:1611.07770 [hep-ph]} \BibitemShut
  {NoStop}%
\bibitem [{\citenamefont {Adler}(1968)}]{Adler:1968tw}%
  \BibitemOpen
  \bibfield  {author} {\bibinfo {author} {\bibfnamefont {S.~L.}\ \bibnamefont
  {Adler}},\ }\href {\doibase 10.1016/0003-4916(68)90278-9} {\bibfield
  {journal} {\bibinfo  {journal} {Annals Phys.}\ }\textbf {\bibinfo {volume}
  {50}},\ \bibinfo {pages} {189} (\bibinfo {year} {1968})}\BibitemShut
  {NoStop}%
\bibitem [{\citenamefont {Bijtebier}(1970)}]{Bijtebier:1970ku}%
  \BibitemOpen
  \bibfield  {author} {\bibinfo {author} {\bibfnamefont {J.}~\bibnamefont
  {Bijtebier}},\ }\href {\doibase 10.1016/0550-3213(70)90512-2} {\bibfield
  {journal} {\bibinfo  {journal} {Nucl. Phys.}\ }\textbf {\bibinfo {volume}
  {B21}},\ \bibinfo {pages} {158} (\bibinfo {year} {1970})}\BibitemShut
  {NoStop}%
\bibitem [{\citenamefont {Llewellyn~Smith}(1972)}]{LlewellynSmith:1971uhs}%
  \BibitemOpen
  \bibfield  {author} {\bibinfo {author} {\bibfnamefont {C.~H.}\ \bibnamefont
  {Llewellyn~Smith}},\ }\href {\doibase 10.1016/0370-1573(72)90010-5}
  {\bibfield  {journal} {\bibinfo  {journal} {Phys. Rept.}\ }\textbf {\bibinfo
  {volume} {3}},\ \bibinfo {pages} {261} (\bibinfo {year} {1972})}\BibitemShut
  {NoStop}%
\bibitem [{\citenamefont {Alevizos}\ \emph {et~al.}(1977)\citenamefont
  {Alevizos}, \citenamefont {Celikel},\ and\ \citenamefont
  {Dombey}}]{Alevizos:1977xf}%
  \BibitemOpen
  \bibfield  {author} {\bibinfo {author} {\bibfnamefont {T.}~\bibnamefont
  {Alevizos}}, \bibinfo {author} {\bibfnamefont {A.}~\bibnamefont {Celikel}}, \
  and\ \bibinfo {author} {\bibfnamefont {N.}~\bibnamefont {Dombey}},\ }\href
  {\doibase 10.1088/0305-4616/3/9/010} {\bibfield  {journal} {\bibinfo
  {journal} {J. Phys.}\ }\textbf {\bibinfo {volume} {G3}},\ \bibinfo {pages}
  {1179} (\bibinfo {year} {1977})}\BibitemShut {NoStop}%
\bibitem [{\citenamefont {Fogli}\ and\ \citenamefont
  {Nardulli}(1979)}]{Fogli:1979cz}%
  \BibitemOpen
  \bibfield  {author} {\bibinfo {author} {\bibfnamefont {G.~L.}\ \bibnamefont
  {Fogli}}\ and\ \bibinfo {author} {\bibfnamefont {G.}~\bibnamefont
  {Nardulli}},\ }\href {\doibase 10.1016/0550-3213(79)90233-5} {\bibfield
  {journal} {\bibinfo  {journal} {Nucl. Phys.}\ }\textbf {\bibinfo {volume}
  {B160}},\ \bibinfo {pages} {116} (\bibinfo {year} {1979})}\BibitemShut
  {NoStop}%
\bibitem [{\citenamefont {Rein}\ and\ \citenamefont
  {Sehgal}(1981)}]{Rein:1980wg}%
  \BibitemOpen
  \bibfield  {author} {\bibinfo {author} {\bibfnamefont {D.}~\bibnamefont
  {Rein}}\ and\ \bibinfo {author} {\bibfnamefont {L.~M.}\ \bibnamefont
  {Sehgal}},\ }\href {\doibase 10.1016/0003-4916(81)90242-6} {\bibfield
  {journal} {\bibinfo  {journal} {Annals Phys.}\ }\textbf {\bibinfo {volume}
  {133}},\ \bibinfo {pages} {79} (\bibinfo {year} {1981})}\BibitemShut
  {NoStop}%
\bibitem [{\citenamefont {Hernandez}\ \emph {et~al.}(2007)\citenamefont
  {Hernandez}, \citenamefont {Nieves},\ and\ \citenamefont
  {Valverde}}]{Hernandez:2007qq}%
  \BibitemOpen
  \bibfield  {author} {\bibinfo {author} {\bibfnamefont {E.}~\bibnamefont
  {Hernandez}}, \bibinfo {author} {\bibfnamefont {J.}~\bibnamefont {Nieves}}, \
  and\ \bibinfo {author} {\bibfnamefont {M.}~\bibnamefont {Valverde}},\ }\href
  {\doibase 10.1103/PhysRevD.76.033005} {\bibfield  {journal} {\bibinfo
  {journal} {Phys. Rev.}\ }\textbf {\bibinfo {volume} {D76}},\ \bibinfo {pages}
  {033005} (\bibinfo {year} {2007})},\ \Eprint
  {http://arxiv.org/abs/hep-ph/0701149} {arXiv:hep-ph/0701149 [hep-ph]}
  \BibitemShut {NoStop}%
\bibitem [{\citenamefont {Leitner}\ \emph {et~al.}(2009)\citenamefont
  {Leitner}, \citenamefont {Buss}, \citenamefont {Alvarez-Ruso},\ and\
  \citenamefont {Mosel}}]{Leitner:2008ue}%
  \BibitemOpen
  \bibfield  {author} {\bibinfo {author} {\bibfnamefont {T.}~\bibnamefont
  {Leitner}}, \bibinfo {author} {\bibfnamefont {O.}~\bibnamefont {Buss}},
  \bibinfo {author} {\bibfnamefont {L.}~\bibnamefont {Alvarez-Ruso}}, \ and\
  \bibinfo {author} {\bibfnamefont {U.}~\bibnamefont {Mosel}},\ }\href
  {\doibase 10.1103/PhysRevC.79.034601} {\bibfield  {journal} {\bibinfo
  {journal} {Phys. Rev.}\ }\textbf {\bibinfo {volume} {C79}},\ \bibinfo {pages}
  {034601} (\bibinfo {year} {2009})},\ \Eprint {http://arxiv.org/abs/0812.0587}
  {arXiv:0812.0587 [nucl-th]} \BibitemShut {NoStop}%
\bibitem [{\citenamefont {Barbero}\ \emph {et~al.}(2008)\citenamefont
  {Barbero}, \citenamefont {Lopez~Castro},\ and\ \citenamefont
  {Mariano}}]{Barbero:2008zza}%
  \BibitemOpen
  \bibfield  {author} {\bibinfo {author} {\bibfnamefont {C.}~\bibnamefont
  {Barbero}}, \bibinfo {author} {\bibfnamefont {G.}~\bibnamefont
  {Lopez~Castro}}, \ and\ \bibinfo {author} {\bibfnamefont {A.}~\bibnamefont
  {Mariano}},\ }\href {\doibase 10.1016/j.physletb.2008.05.011} {\bibfield
  {journal} {\bibinfo  {journal} {Phys. Lett.}\ }\textbf {\bibinfo {volume}
  {B664}},\ \bibinfo {pages} {70} (\bibinfo {year} {2008})}\BibitemShut
  {NoStop}%
\bibitem [{\citenamefont {Sato}\ and\ \citenamefont {Lee}(2009)}]{Sato:2009de}%
  \BibitemOpen
  \bibfield  {author} {\bibinfo {author} {\bibfnamefont {T.}~\bibnamefont
  {Sato}}\ and\ \bibinfo {author} {\bibfnamefont {T.~S.~H.}\ \bibnamefont
  {Lee}},\ }\href {\doibase 10.1088/0954-3899/36/7/073001} {\bibfield
  {journal} {\bibinfo  {journal} {J. Phys.}\ }\textbf {\bibinfo {volume}
  {G36}},\ \bibinfo {pages} {073001} (\bibinfo {year} {2009})},\ \Eprint
  {http://arxiv.org/abs/0902.3653} {arXiv:0902.3653 [nucl-th]} \BibitemShut
  {NoStop}%
\bibitem [{\citenamefont {Serot}\ and\ \citenamefont
  {Zhang}(2012)}]{Serot:2012rd}%
  \BibitemOpen
  \bibfield  {author} {\bibinfo {author} {\bibfnamefont {B.~D.}\ \bibnamefont
  {Serot}}\ and\ \bibinfo {author} {\bibfnamefont {X.}~\bibnamefont {Zhang}},\
  }\href {\doibase 10.1103/PhysRevC.86.015501} {\bibfield  {journal} {\bibinfo
  {journal} {Phys. Rev.}\ }\textbf {\bibinfo {volume} {C86}},\ \bibinfo {pages}
  {015501} (\bibinfo {year} {2012})},\ \Eprint {http://arxiv.org/abs/1206.3812}
  {arXiv:1206.3812 [nucl-th]} \BibitemShut {NoStop}%
\bibitem [{\citenamefont {Graczyk}\ \emph {et~al.}(2014)\citenamefont
  {Graczyk}, \citenamefont {Zmuda},\ and\ \citenamefont
  {Sobczyk}}]{Graczyk:2014dpa}%
  \BibitemOpen
  \bibfield  {author} {\bibinfo {author} {\bibfnamefont {K.~M.}\ \bibnamefont
  {Graczyk}}, \bibinfo {author} {\bibfnamefont {J.}~\bibnamefont {Zmuda}}, \
  and\ \bibinfo {author} {\bibfnamefont {J.~T.}\ \bibnamefont {Sobczyk}},\
  }\href {\doibase 10.1103/PhysRevD.90.093001} {\bibfield  {journal} {\bibinfo
  {journal} {Phys. Rev.}\ }\textbf {\bibinfo {volume} {D90}},\ \bibinfo {pages}
  {093001} (\bibinfo {year} {2014})},\ \Eprint {http://arxiv.org/abs/1407.5445}
  {arXiv:1407.5445 [hep-ph]} \BibitemShut {NoStop}%
\bibitem [{\citenamefont {Nakamura}\ \emph {et~al.}(2015)\citenamefont
  {Nakamura}, \citenamefont {Kamano},\ and\ \citenamefont
  {Sato}}]{Nakamura:2015rta}%
  \BibitemOpen
  \bibfield  {author} {\bibinfo {author} {\bibfnamefont {S.~X.}\ \bibnamefont
  {Nakamura}}, \bibinfo {author} {\bibfnamefont {H.}~\bibnamefont {Kamano}}, \
  and\ \bibinfo {author} {\bibfnamefont {T.}~\bibnamefont {Sato}},\ }\href
  {\doibase 10.1103/PhysRevD.92.074024} {\bibfield  {journal} {\bibinfo
  {journal} {Phys. Rev.}\ }\textbf {\bibinfo {volume} {D92}},\ \bibinfo {pages}
  {074024} (\bibinfo {year} {2015})},\ \Eprint
  {http://arxiv.org/abs/1506.03403} {arXiv:1506.03403 [hep-ph]} \BibitemShut
  {NoStop}%
\bibitem [{\citenamefont {Rafi~Alam}\ \emph {et~al.}(2016)\citenamefont
  {Rafi~Alam}, \citenamefont {Sajjad~Athar}, \citenamefont {Chauhan},\ and\
  \citenamefont {Singh}}]{Alam:2015gaa}%
  \BibitemOpen
  \bibfield  {author} {\bibinfo {author} {\bibfnamefont {M.}~\bibnamefont
  {Rafi~Alam}}, \bibinfo {author} {\bibfnamefont {M.}~\bibnamefont
  {Sajjad~Athar}}, \bibinfo {author} {\bibfnamefont {S.}~\bibnamefont
  {Chauhan}}, \ and\ \bibinfo {author} {\bibfnamefont {S.~K.}\ \bibnamefont
  {Singh}},\ }\href {\doibase 10.1142/S0218301316500105} {\bibfield  {journal}
  {\bibinfo  {journal} {Int. J. Mod. Phys.}\ }\textbf {\bibinfo {volume}
  {E25}},\ \bibinfo {pages} {1650010} (\bibinfo {year} {2016})},\ \Eprint
  {http://arxiv.org/abs/1509.08622} {arXiv:1509.08622 [hep-ph]} \BibitemShut
  {NoStop}%
\bibitem [{\citenamefont {González-Jiménez}\ \emph
  {et~al.}(2017)\citenamefont {González-Jiménez}, \citenamefont {Jachowicz},
  \citenamefont {Niewczas}, \citenamefont {Nys}, \citenamefont {Pandey},
  \citenamefont {Van~Cuyck},\ and\ \citenamefont
  {Van~Dessel}}]{Gonzalez-Jimenez:2016qqq}%
  \BibitemOpen
  \bibfield  {author} {\bibinfo {author} {\bibfnamefont {R.}~\bibnamefont
  {González-Jiménez}}, \bibinfo {author} {\bibfnamefont {N.}~\bibnamefont
  {Jachowicz}}, \bibinfo {author} {\bibfnamefont {K.}~\bibnamefont {Niewczas}},
  \bibinfo {author} {\bibfnamefont {J.}~\bibnamefont {Nys}}, \bibinfo {author}
  {\bibfnamefont {V.}~\bibnamefont {Pandey}}, \bibinfo {author} {\bibfnamefont
  {T.}~\bibnamefont {Van~Cuyck}}, \ and\ \bibinfo {author} {\bibfnamefont
  {N.}~\bibnamefont {Van~Dessel}},\ }\href {\doibase
  10.1103/PhysRevD.95.113007} {\bibfield  {journal} {\bibinfo  {journal} {Phys.
  Rev.}\ }\textbf {\bibinfo {volume} {D95}},\ \bibinfo {pages} {113007}
  (\bibinfo {year} {2017})},\ \Eprint {http://arxiv.org/abs/1612.05511}
  {arXiv:1612.05511 [nucl-th]} \BibitemShut {NoStop}%
\bibitem [{\citenamefont {Kabirnezhad}(2018)}]{Kabirnezhad:2017jmf}%
  \BibitemOpen
  \bibfield  {author} {\bibinfo {author} {\bibfnamefont {M.}~\bibnamefont
  {Kabirnezhad}},\ }\bibfield  {booktitle} {\emph {\bibinfo {booktitle} {{11th
  International Workshop on Neutrino-Nucleus Scattering in the Few GeV Region
  (NuInt17) Toronto, Ontario, Canada, June 25-30, 2017}}},\ }\href {\doibase
  10.1103/PhysRevD.97.013002} {\bibfield  {journal} {\bibinfo  {journal} {Phys.
  Rev.}\ }\textbf {\bibinfo {volume} {D97}},\ \bibinfo {pages} {013002}
  (\bibinfo {year} {2018})},\ \Eprint {http://arxiv.org/abs/1711.02403}
  {arXiv:1711.02403 [hep-ph]} \BibitemShut {NoStop}%
\bibitem [{\citenamefont {Fogli}\ and\ \citenamefont
  {Nardulli}(1980)}]{Fogli:1979qj}%
  \BibitemOpen
  \bibfield  {author} {\bibinfo {author} {\bibfnamefont {G.~L.}\ \bibnamefont
  {Fogli}}\ and\ \bibinfo {author} {\bibfnamefont {G.}~\bibnamefont
  {Nardulli}},\ }\href {\doibase 10.1016/0550-3213(80)90312-0} {\bibfield
  {journal} {\bibinfo  {journal} {Nucl. Phys.}\ }\textbf {\bibinfo {volume}
  {B165}},\ \bibinfo {pages} {162} (\bibinfo {year} {1980})}\BibitemShut
  {NoStop}%
\bibitem [{\citenamefont {Leitner}\ \emph {et~al.}(2006)\citenamefont
  {Leitner}, \citenamefont {Alvarez-Ruso},\ and\ \citenamefont
  {Mosel}}]{Leitner:2006sp}%
  \BibitemOpen
  \bibfield  {author} {\bibinfo {author} {\bibfnamefont {T.}~\bibnamefont
  {Leitner}}, \bibinfo {author} {\bibfnamefont {L.}~\bibnamefont
  {Alvarez-Ruso}}, \ and\ \bibinfo {author} {\bibfnamefont {U.}~\bibnamefont
  {Mosel}},\ }\href {\doibase 10.1103/PhysRevC.74.065502} {\bibfield  {journal}
  {\bibinfo  {journal} {Phys. Rev.}\ }\textbf {\bibinfo {volume} {C74}},\
  \bibinfo {pages} {065502} (\bibinfo {year} {2006})},\ \Eprint
  {http://arxiv.org/abs/nucl-th/0606058} {arXiv:nucl-th/0606058 [nucl-th]}
  \BibitemShut {NoStop}%
\bibitem [{\citenamefont {Yao}\ \emph {et~al.}(2018)\citenamefont {Yao},
  \citenamefont {Alvarez-Ruso}, \citenamefont {Hiller~Blin},\ and\
  \citenamefont {Vicente~Vacas}}]{Yao:2018pzc}%
  \BibitemOpen
  \bibfield  {author} {\bibinfo {author} {\bibfnamefont {D.-L.}\ \bibnamefont
  {Yao}}, \bibinfo {author} {\bibfnamefont {L.}~\bibnamefont {Alvarez-Ruso}},
  \bibinfo {author} {\bibfnamefont {A.~N.}\ \bibnamefont {Hiller~Blin}}, \ and\
  \bibinfo {author} {\bibfnamefont {M.~J.}\ \bibnamefont {Vicente~Vacas}},\
  }\href {\doibase 10.1103/PhysRevD.98.076004} {\bibfield  {journal} {\bibinfo
  {journal} {Phys. Rev.}\ }\textbf {\bibinfo {volume} {D98}},\ \bibinfo {pages}
  {076004} (\bibinfo {year} {2018})},\ \Eprint
  {http://arxiv.org/abs/1806.09364} {arXiv:1806.09364 [hep-ph]} \BibitemShut
  {NoStop}%
\bibitem [{\citenamefont {Weinberg}(1979)}]{Weinberg:1978kz}%
  \BibitemOpen
  \bibfield  {author} {\bibinfo {author} {\bibfnamefont {S.}~\bibnamefont
  {Weinberg}},\ }\href {\doibase 10.1016/0378-4371(79)90223-1} {\bibfield
  {journal} {\bibinfo  {journal} {Physica}\ }\textbf {\bibinfo {volume}
  {A96}},\ \bibinfo {pages} {327} (\bibinfo {year} {1979})}\BibitemShut
  {NoStop}%
\bibitem [{\citenamefont {Gasser}\ and\ \citenamefont
  {Leutwyler}(1984)}]{Gasser:1983yg}%
  \BibitemOpen
  \bibfield  {author} {\bibinfo {author} {\bibfnamefont {J.}~\bibnamefont
  {Gasser}}\ and\ \bibinfo {author} {\bibfnamefont {H.}~\bibnamefont
  {Leutwyler}},\ }\href {\doibase 10.1016/0003-4916(84)90242-2} {\bibfield
  {journal} {\bibinfo  {journal} {Annals Phys.}\ }\textbf {\bibinfo {volume}
  {158}},\ \bibinfo {pages} {142} (\bibinfo {year} {1984})}\BibitemShut
  {NoStop}%
\bibitem [{\citenamefont {Gasser}\ and\ \citenamefont
  {Leutwyler}(1985)}]{Gasser:1984gg}%
  \BibitemOpen
  \bibfield  {author} {\bibinfo {author} {\bibfnamefont {J.}~\bibnamefont
  {Gasser}}\ and\ \bibinfo {author} {\bibfnamefont {H.}~\bibnamefont
  {Leutwyler}},\ }\href {\doibase 10.1016/0550-3213(85)90492-4} {\bibfield
  {journal} {\bibinfo  {journal} {Nucl. Phys.}\ }\textbf {\bibinfo {volume}
  {B250}},\ \bibinfo {pages} {465} (\bibinfo {year} {1985})}\BibitemShut
  {NoStop}%
\bibitem [{\citenamefont {Bernard}\ \emph {et~al.}(1995)\citenamefont
  {Bernard}, \citenamefont {Kaiser},\ and\ \citenamefont
  {Mei{\ss}ner}}]{Bernard:1995dp}%
  \BibitemOpen
  \bibfield  {author} {\bibinfo {author} {\bibfnamefont {V.}~\bibnamefont
  {Bernard}}, \bibinfo {author} {\bibfnamefont {N.}~\bibnamefont {Kaiser}}, \
  and\ \bibinfo {author} {\bibfnamefont {U.-G.}\ \bibnamefont {Mei{\ss}ner}},\
  }\href {\doibase 10.1142/S0218301395000092} {\bibfield  {journal} {\bibinfo
  {journal} {Int. J. Mod. Phys.}\ }\textbf {\bibinfo {volume} {E4}},\ \bibinfo
  {pages} {193} (\bibinfo {year} {1995})},\ \Eprint
  {http://arxiv.org/abs/hep-ph/9501384} {arXiv:hep-ph/9501384 [hep-ph]}
  \BibitemShut {NoStop}%
\bibitem [{\citenamefont {Scherer}\ and\ \citenamefont
  {Schindler}(2012)}]{Scherer:2012xha}%
  \BibitemOpen
  \bibfield  {author} {\bibinfo {author} {\bibfnamefont {S.}~\bibnamefont
  {Scherer}}\ and\ \bibinfo {author} {\bibfnamefont {M.~R.}\ \bibnamefont
  {Schindler}},\ }\href {\doibase 10.1007/978-3-642-19254-8} {\bibfield
  {journal} {\bibinfo  {journal} {Lect. Notes Phys.}\ }\textbf {\bibinfo
  {volume} {830}},\ \bibinfo {pages} {pp.1} (\bibinfo {year}
  {2012})}\BibitemShut {NoStop}%
\bibitem [{\citenamefont {Fuchs}\ \emph {et~al.}(2003)\citenamefont {Fuchs},
  \citenamefont {Gegelia}, \citenamefont {Japaridze},\ and\ \citenamefont
  {Scherer}}]{Fuchs:2003qc}%
  \BibitemOpen
  \bibfield  {author} {\bibinfo {author} {\bibfnamefont {T.}~\bibnamefont
  {Fuchs}}, \bibinfo {author} {\bibfnamefont {J.}~\bibnamefont {Gegelia}},
  \bibinfo {author} {\bibfnamefont {G.}~\bibnamefont {Japaridze}}, \ and\
  \bibinfo {author} {\bibfnamefont {S.}~\bibnamefont {Scherer}},\ }\href
  {\doibase 10.1103/PhysRevD.68.056005} {\bibfield  {journal} {\bibinfo
  {journal} {Phys. Rev.}\ }\textbf {\bibinfo {volume} {D68}},\ \bibinfo {pages}
  {056005} (\bibinfo {year} {2003})},\ \Eprint
  {http://arxiv.org/abs/hep-ph/0302117} {arXiv:hep-ph/0302117 [hep-ph]}
  \BibitemShut {NoStop}%
\bibitem [{\citenamefont {Gasser}\ \emph {et~al.}(1988)\citenamefont {Gasser},
  \citenamefont {Sainio},\ and\ \citenamefont {Svarc}}]{Gasser:1987rb}%
  \BibitemOpen
  \bibfield  {author} {\bibinfo {author} {\bibfnamefont {J.}~\bibnamefont
  {Gasser}}, \bibinfo {author} {\bibfnamefont {M.~E.}\ \bibnamefont {Sainio}},
  \ and\ \bibinfo {author} {\bibfnamefont {A.}~\bibnamefont {Svarc}},\ }\href
  {\doibase 10.1016/0550-3213(88)90108-3} {\bibfield  {journal} {\bibinfo
  {journal} {Nucl. Phys.}\ }\textbf {\bibinfo {volume} {B307}},\ \bibinfo
  {pages} {779} (\bibinfo {year} {1988})}\BibitemShut {NoStop}%
\bibitem [{\citenamefont {Geng}\ \emph {et~al.}(2008)\citenamefont {Geng},
  \citenamefont {Martin~Camalich}, \citenamefont {Alvarez-Ruso},\ and\
  \citenamefont {Vicente~Vacas}}]{Geng:2008mf}%
  \BibitemOpen
  \bibfield  {author} {\bibinfo {author} {\bibfnamefont {L.~S.}\ \bibnamefont
  {Geng}}, \bibinfo {author} {\bibfnamefont {J.}~\bibnamefont
  {Martin~Camalich}}, \bibinfo {author} {\bibfnamefont {L.}~\bibnamefont
  {Alvarez-Ruso}}, \ and\ \bibinfo {author} {\bibfnamefont {M.~J.}\
  \bibnamefont {Vicente~Vacas}},\ }\href {\doibase
  10.1103/PhysRevLett.101.222002} {\bibfield  {journal} {\bibinfo  {journal}
  {Phys. Rev. Lett.}\ }\textbf {\bibinfo {volume} {101}},\ \bibinfo {pages}
  {222002} (\bibinfo {year} {2008})},\ \Eprint {http://arxiv.org/abs/0805.1419}
  {arXiv:0805.1419 [hep-ph]} \BibitemShut {NoStop}%
\bibitem [{\citenamefont {Alarcon}\ \emph {et~al.}(2013)\citenamefont
  {Alarcon}, \citenamefont {Martin~Camalich},\ and\ \citenamefont
  {Oller}}]{Alarcon:2012kn}%
  \BibitemOpen
  \bibfield  {author} {\bibinfo {author} {\bibfnamefont {J.~M.}\ \bibnamefont
  {Alarcon}}, \bibinfo {author} {\bibfnamefont {J.}~\bibnamefont
  {Martin~Camalich}}, \ and\ \bibinfo {author} {\bibfnamefont {J.~A.}\
  \bibnamefont {Oller}},\ }\href {\doibase 10.1016/j.aop.2013.06.001}
  {\bibfield  {journal} {\bibinfo  {journal} {Annals Phys.}\ }\textbf {\bibinfo
  {volume} {336}},\ \bibinfo {pages} {413} (\bibinfo {year} {2013})},\ \Eprint
  {http://arxiv.org/abs/1210.4450} {arXiv:1210.4450 [hep-ph]} \BibitemShut
  {NoStop}%
\bibitem [{\citenamefont {Chen}\ \emph {et~al.}(2013)\citenamefont {Chen},
  \citenamefont {Yao},\ and\ \citenamefont {Zheng}}]{Chen:2012nx}%
  \BibitemOpen
  \bibfield  {author} {\bibinfo {author} {\bibfnamefont {Y.-H.}\ \bibnamefont
  {Chen}}, \bibinfo {author} {\bibfnamefont {D.-L.}\ \bibnamefont {Yao}}, \
  and\ \bibinfo {author} {\bibfnamefont {H.~Q.}\ \bibnamefont {Zheng}},\ }\href
  {\doibase 10.1103/PhysRevD.87.054019} {\bibfield  {journal} {\bibinfo
  {journal} {Phys. Rev.}\ }\textbf {\bibinfo {volume} {D87}},\ \bibinfo {pages}
  {054019} (\bibinfo {year} {2013})},\ \Eprint {http://arxiv.org/abs/1212.1893}
  {arXiv:1212.1893 [hep-ph]} \BibitemShut {NoStop}%
\bibitem [{\citenamefont {Hilt}\ \emph
  {et~al.}(2013{\natexlab{a}})\citenamefont {Hilt}, \citenamefont {Lehnhart},
  \citenamefont {Scherer},\ and\ \citenamefont {Tiator}}]{Hilt:2013fda}%
  \BibitemOpen
  \bibfield  {author} {\bibinfo {author} {\bibfnamefont {M.}~\bibnamefont
  {Hilt}}, \bibinfo {author} {\bibfnamefont {B.~C.}\ \bibnamefont {Lehnhart}},
  \bibinfo {author} {\bibfnamefont {S.}~\bibnamefont {Scherer}}, \ and\
  \bibinfo {author} {\bibfnamefont {L.}~\bibnamefont {Tiator}},\ }\href
  {\doibase 10.1103/PhysRevC.88.055207} {\bibfield  {journal} {\bibinfo
  {journal} {Phys. Rev.}\ }\textbf {\bibinfo {volume} {C88}},\ \bibinfo {pages}
  {055207} (\bibinfo {year} {2013}{\natexlab{a}})},\ \Eprint
  {http://arxiv.org/abs/1309.3385} {arXiv:1309.3385 [nucl-th]} \BibitemShut
  {NoStop}%
\bibitem [{\citenamefont {Yao}\ \emph {et~al.}(2016)\citenamefont {Yao},
  \citenamefont {Siemens}, \citenamefont {Bernard}, \citenamefont {Epelbaum},
  \citenamefont {Gasparyan}, \citenamefont {Gegelia}, \citenamefont {Krebs},\
  and\ \citenamefont {Mei{\ss}ner}}]{Yao:2016vbz}%
  \BibitemOpen
  \bibfield  {author} {\bibinfo {author} {\bibfnamefont {D.-L.}\ \bibnamefont
  {Yao}}, \bibinfo {author} {\bibfnamefont {D.}~\bibnamefont {Siemens}},
  \bibinfo {author} {\bibfnamefont {V.}~\bibnamefont {Bernard}}, \bibinfo
  {author} {\bibfnamefont {E.}~\bibnamefont {Epelbaum}}, \bibinfo {author}
  {\bibfnamefont {A.~M.}\ \bibnamefont {Gasparyan}}, \bibinfo {author}
  {\bibfnamefont {J.}~\bibnamefont {Gegelia}}, \bibinfo {author} {\bibfnamefont
  {H.}~\bibnamefont {Krebs}}, \ and\ \bibinfo {author} {\bibfnamefont {U.-G.}\
  \bibnamefont {Mei{\ss}ner}},\ }\href {\doibase 10.1007/JHEP05(2016)038}
  {\bibfield  {journal} {\bibinfo  {journal} {JHEP}\ }\textbf {\bibinfo
  {volume} {05}},\ \bibinfo {pages} {038} (\bibinfo {year} {2016})},\ \Eprint
  {http://arxiv.org/abs/1603.03638} {arXiv:1603.03638 [hep-ph]} \BibitemShut
  {NoStop}%
\bibitem [{\citenamefont {Hiller~Blin}\ \emph {et~al.}(2016)\citenamefont
  {Hiller~Blin}, \citenamefont {Ledwig},\ and\ \citenamefont
  {Vicente~Vacas}}]{Blin:2016itn}%
  \BibitemOpen
  \bibfield  {author} {\bibinfo {author} {\bibfnamefont {A.~N.}\ \bibnamefont
  {Hiller~Blin}}, \bibinfo {author} {\bibfnamefont {T.}~\bibnamefont {Ledwig}},
  \ and\ \bibinfo {author} {\bibfnamefont {M.~J.}\ \bibnamefont
  {Vicente~Vacas}},\ }\href {\doibase 10.1103/PhysRevD.93.094018} {\bibfield
  {journal} {\bibinfo  {journal} {Phys. Rev.}\ }\textbf {\bibinfo {volume}
  {D93}},\ \bibinfo {pages} {094018} (\bibinfo {year} {2016})},\ \Eprint
  {http://arxiv.org/abs/1602.08967} {arXiv:1602.08967 [hep-ph]} \BibitemShut
  {NoStop}%
\bibitem [{\citenamefont {Siemens}\ \emph {et~al.}(2017)\citenamefont
  {Siemens}, \citenamefont {Bernard}, \citenamefont {Epelbaum}, \citenamefont
  {Gasparyan}, \citenamefont {Krebs},\ and\ \citenamefont
  {Mei{\ss}ner}}]{Siemens:2017opr}%
  \BibitemOpen
  \bibfield  {author} {\bibinfo {author} {\bibfnamefont {D.}~\bibnamefont
  {Siemens}}, \bibinfo {author} {\bibfnamefont {V.}~\bibnamefont {Bernard}},
  \bibinfo {author} {\bibfnamefont {E.}~\bibnamefont {Epelbaum}}, \bibinfo
  {author} {\bibfnamefont {A.~M.}\ \bibnamefont {Gasparyan}}, \bibinfo {author}
  {\bibfnamefont {H.}~\bibnamefont {Krebs}}, \ and\ \bibinfo {author}
  {\bibfnamefont {U.-G.}\ \bibnamefont {Mei{\ss}ner}},\ }\href {\doibase
  10.1103/PhysRevC.96.055205} {\bibfield  {journal} {\bibinfo  {journal} {Phys.
  Rev.}\ }\textbf {\bibinfo {volume} {C96}},\ \bibinfo {pages} {055205}
  (\bibinfo {year} {2017})},\ \Eprint {http://arxiv.org/abs/1704.08988}
  {arXiv:1704.08988 [nucl-th]} \BibitemShut {NoStop}%
\bibitem [{\citenamefont {Yao}\ \emph {et~al.}(2017)\citenamefont {Yao},
  \citenamefont {Alvarez-Ruso},\ and\ \citenamefont
  {Vicente-Vacas}}]{Yao:2017fym}%
  \BibitemOpen
  \bibfield  {author} {\bibinfo {author} {\bibfnamefont {D.-L.}\ \bibnamefont
  {Yao}}, \bibinfo {author} {\bibfnamefont {L.}~\bibnamefont {Alvarez-Ruso}}, \
  and\ \bibinfo {author} {\bibfnamefont {M.~J.}\ \bibnamefont
  {Vicente-Vacas}},\ }\href {\doibase 10.1103/PhysRevD.96.116022} {\bibfield
  {journal} {\bibinfo  {journal} {Phys. Rev.}\ }\textbf {\bibinfo {volume}
  {D96}},\ \bibinfo {pages} {116022} (\bibinfo {year} {2017})},\ \Eprint
  {http://arxiv.org/abs/1708.08776} {arXiv:1708.08776 [hep-ph]} \BibitemShut
  {NoStop}%
\bibitem [{\citenamefont {Epelbaum}\ \emph
  {et~al.}(2015{\natexlab{a}})\citenamefont {Epelbaum}, \citenamefont
  {Gegelia}, \citenamefont {Mei{\ss}ner},\ and\ \citenamefont
  {Yao}}]{Epelbaum:2015vea}%
  \BibitemOpen
  \bibfield  {author} {\bibinfo {author} {\bibfnamefont {E.}~\bibnamefont
  {Epelbaum}}, \bibinfo {author} {\bibfnamefont {J.}~\bibnamefont {Gegelia}},
  \bibinfo {author} {\bibfnamefont {U.-G.}\ \bibnamefont {Mei{\ss}ner}}, \ and\
  \bibinfo {author} {\bibfnamefont {D.-L.}\ \bibnamefont {Yao}},\ }\href
  {\doibase 10.1140/epjc/s10052-015-3728-7} {\bibfield  {journal} {\bibinfo
  {journal} {Eur. Phys. J.}\ }\textbf {\bibinfo {volume} {C75}},\ \bibinfo
  {pages} {499} (\bibinfo {year} {2015}{\natexlab{a}})},\ \Eprint
  {http://arxiv.org/abs/1510.02388} {arXiv:1510.02388 [hep-ph]} \BibitemShut
  {NoStop}%
\bibitem [{\citenamefont {Fernandez-Ramirez}\ and\ \citenamefont
  {Bernstein}(2013)}]{FernandezRamirez:2012nw}%
  \BibitemOpen
  \bibfield  {author} {\bibinfo {author} {\bibfnamefont {C.}~\bibnamefont
  {Fernandez-Ramirez}}\ and\ \bibinfo {author} {\bibfnamefont {A.~M.}\
  \bibnamefont {Bernstein}},\ }\href {\doibase 10.1016/j.physletb.2013.06.020}
  {\bibfield  {journal} {\bibinfo  {journal} {Phys. Lett.}\ }\textbf {\bibinfo
  {volume} {B724}},\ \bibinfo {pages} {253} (\bibinfo {year} {2013})},\ \Eprint
  {http://arxiv.org/abs/1212.3237} {arXiv:1212.3237 [nucl-th]} \BibitemShut
  {NoStop}%
\bibitem [{\citenamefont {Hilt}\ \emph
  {et~al.}(2013{\natexlab{b}})\citenamefont {Hilt}, \citenamefont {Scherer},\
  and\ \citenamefont {Tiator}}]{Hilt:2013uf}%
  \BibitemOpen
  \bibfield  {author} {\bibinfo {author} {\bibfnamefont {M.}~\bibnamefont
  {Hilt}}, \bibinfo {author} {\bibfnamefont {S.}~\bibnamefont {Scherer}}, \
  and\ \bibinfo {author} {\bibfnamefont {L.}~\bibnamefont {Tiator}},\ }\href
  {\doibase 10.1103/PhysRevC.87.045204} {\bibfield  {journal} {\bibinfo
  {journal} {Phys. Rev.}\ }\textbf {\bibinfo {volume} {C87}},\ \bibinfo {pages}
  {045204} (\bibinfo {year} {2013}{\natexlab{b}})},\ \Eprint
  {http://arxiv.org/abs/1301.5576} {arXiv:1301.5576 [nucl-th]} \BibitemShut
  {NoStop}%
\bibitem [{\citenamefont {Bernard}\ \emph {et~al.}(1994)\citenamefont
  {Bernard}, \citenamefont {Kaiser},\ and\ \citenamefont
  {Mei{\ss}ner}}]{Bernard:1993xh}%
  \BibitemOpen
  \bibfield  {author} {\bibinfo {author} {\bibfnamefont {V.}~\bibnamefont
  {Bernard}}, \bibinfo {author} {\bibfnamefont {N.}~\bibnamefont {Kaiser}}, \
  and\ \bibinfo {author} {\bibfnamefont {U.~G.}\ \bibnamefont {Mei{\ss}ner}},\
  }\href {\doibase 10.1016/0370-2693(94)90954-7} {\bibfield  {journal}
  {\bibinfo  {journal} {Phys. Lett.}\ }\textbf {\bibinfo {volume} {B331}},\
  \bibinfo {pages} {137} (\bibinfo {year} {1994})},\ \Eprint
  {http://arxiv.org/abs/hep-ph/9312307} {arXiv:hep-ph/9312307 [hep-ph]}
  \BibitemShut {NoStop}%
\bibitem [{\citenamefont {Pascalutsa}\ and\ \citenamefont
  {Phillips}(2003)}]{Pascalutsa:2002pi}%
  \BibitemOpen
  \bibfield  {author} {\bibinfo {author} {\bibfnamefont {V.}~\bibnamefont
  {Pascalutsa}}\ and\ \bibinfo {author} {\bibfnamefont {D.~R.}\ \bibnamefont
  {Phillips}},\ }\href {\doibase 10.1103/PhysRevC.67.055202} {\bibfield
  {journal} {\bibinfo  {journal} {Phys. Rev.}\ }\textbf {\bibinfo {volume}
  {C67}},\ \bibinfo {pages} {055202} (\bibinfo {year} {2003})},\ \Eprint
  {http://arxiv.org/abs/nucl-th/0212024} {arXiv:nucl-th/0212024 [nucl-th]}
  \BibitemShut {NoStop}%
\bibitem [{\citenamefont {Fettes}\ \emph {et~al.}(2000)\citenamefont {Fettes},
  \citenamefont {Mei{\ss}ner}, \citenamefont {Mojzis},\ and\ \citenamefont
  {Steininger}}]{Fettes:2000gb}%
  \BibitemOpen
  \bibfield  {author} {\bibinfo {author} {\bibfnamefont {N.}~\bibnamefont
  {Fettes}}, \bibinfo {author} {\bibfnamefont {U.-G.}\ \bibnamefont
  {Mei{\ss}ner}}, \bibinfo {author} {\bibfnamefont {M.}~\bibnamefont {Mojzis}},
  \ and\ \bibinfo {author} {\bibfnamefont {S.}~\bibnamefont {Steininger}},\
  }\href {\doibase 10.1006/aphy.2000.6059} {\bibfield  {journal} {\bibinfo
  {journal} {Annals Phys.}\ }\textbf {\bibinfo {volume} {283}},\ \bibinfo
  {pages} {273} (\bibinfo {year} {2000})},\ \bibinfo {note} {[Erratum: Annals
  Phys.288,249(2001)]},\ \Eprint {http://arxiv.org/abs/hep-ph/0001308}
  {arXiv:hep-ph/0001308 [hep-ph]} \BibitemShut {NoStop}%
\bibitem [{\citenamefont {Bauer}\ \emph {et~al.}(2012)\citenamefont {Bauer},
  \citenamefont {Bernauer},\ and\ \citenamefont {Scherer}}]{Bauer:2012pv}%
  \BibitemOpen
  \bibfield  {author} {\bibinfo {author} {\bibfnamefont {T.}~\bibnamefont
  {Bauer}}, \bibinfo {author} {\bibfnamefont {J.~C.}\ \bibnamefont {Bernauer}},
  \ and\ \bibinfo {author} {\bibfnamefont {S.}~\bibnamefont {Scherer}},\ }\href
  {\doibase 10.1103/PhysRevC.86.065206} {\bibfield  {journal} {\bibinfo
  {journal} {Phys. Rev.}\ }\textbf {\bibinfo {volume} {C86}},\ \bibinfo {pages}
  {065206} (\bibinfo {year} {2012})},\ \Eprint {http://arxiv.org/abs/1209.3872}
  {arXiv:1209.3872 [nucl-th]} \BibitemShut {NoStop}%
\bibitem [{\citenamefont {Patrignani}\ \emph {et~al.}(2016)\citenamefont
  {Patrignani} \emph {et~al.}}]{Patrignani:2016xqp}%
  \BibitemOpen
  \bibfield  {author} {\bibinfo {author} {\bibfnamefont {C.}~\bibnamefont
  {Patrignani}} \emph {et~al.} (\bibinfo {collaboration} {Particle Data
  Group}),\ }\href {\doibase 10.1088/1674-1137/40/10/100001} {\bibfield
  {journal} {\bibinfo  {journal} {Chin. Phys.}\ }\textbf {\bibinfo {volume}
  {C40}},\ \bibinfo {pages} {100001} (\bibinfo {year} {2016})}\BibitemShut
  {NoStop}%
\bibitem [{\citenamefont {Fuchs}\ \emph {et~al.}(2004)\citenamefont {Fuchs},
  \citenamefont {Gegelia},\ and\ \citenamefont {Scherer}}]{Fuchs:2003ir}%
  \BibitemOpen
  \bibfield  {author} {\bibinfo {author} {\bibfnamefont {T.}~\bibnamefont
  {Fuchs}}, \bibinfo {author} {\bibfnamefont {J.}~\bibnamefont {Gegelia}}, \
  and\ \bibinfo {author} {\bibfnamefont {S.}~\bibnamefont {Scherer}},\ }\href
  {\doibase 10.1088/0954-3899/30/10/008} {\bibfield  {journal} {\bibinfo
  {journal} {J. Phys.}\ }\textbf {\bibinfo {volume} {G30}},\ \bibinfo {pages}
  {1407} (\bibinfo {year} {2004})},\ \Eprint
  {http://arxiv.org/abs/nucl-th/0305070} {arXiv:nucl-th/0305070 [nucl-th]}
  \BibitemShut {NoStop}%
\bibitem [{\citenamefont {Juszczak}\ \emph {et~al.}(2006)\citenamefont
  {Juszczak}, \citenamefont {Nowak},\ and\ \citenamefont
  {Sobczyk}}]{Juszczak:2005zs}%
  \BibitemOpen
  \bibfield  {author} {\bibinfo {author} {\bibfnamefont {C.}~\bibnamefont
  {Juszczak}}, \bibinfo {author} {\bibfnamefont {J.~A.}\ \bibnamefont {Nowak}},
  \ and\ \bibinfo {author} {\bibfnamefont {J.~T.}\ \bibnamefont {Sobczyk}},\
  }\bibfield  {booktitle} {\emph {\bibinfo {booktitle} {{NuInt05, proceedings
  of the 4th International Workshop on Neutrino-Nucleus Interactions in the
  Few-GeV Region, Okayama, Japan, 26-29 September 2005}}},\ }\href {\doibase
  10.1016/j.nuclphysbps.2006.08.069} {\bibfield  {journal} {\bibinfo  {journal}
  {Nucl. Phys. Proc. Suppl.}\ }\textbf {\bibinfo {volume} {159}},\ \bibinfo
  {pages} {211} (\bibinfo {year} {2006})},\ \bibinfo {note} {[,211(2005)]},\
  \Eprint {http://arxiv.org/abs/hep-ph/0512365} {arXiv:hep-ph/0512365 [hep-ph]}
  \BibitemShut {NoStop}%
\bibitem [{\citenamefont {Andreopoulos}\ \emph {et~al.}(2010)\citenamefont
  {Andreopoulos} \emph {et~al.}}]{Andreopoulos:2009rq}%
  \BibitemOpen
  \bibfield  {author} {\bibinfo {author} {\bibfnamefont {C.}~\bibnamefont
  {Andreopoulos}} \emph {et~al.},\ }\href {\doibase 10.1016/j.nima.2009.12.009}
  {\bibfield  {journal} {\bibinfo  {journal} {Nucl. Instrum. Meth.}\ }\textbf
  {\bibinfo {volume} {A614}},\ \bibinfo {pages} {87} (\bibinfo {year}
  {2010})},\ \Eprint {http://arxiv.org/abs/0905.2517} {arXiv:0905.2517
  [hep-ph]} \BibitemShut {NoStop}%
\bibitem [{\citenamefont {Epelbaum}\ \emph
  {et~al.}(2015{\natexlab{b}})\citenamefont {Epelbaum}, \citenamefont {Krebs},\
  and\ \citenamefont {Meißner}}]{Epelbaum:2014efa}%
  \BibitemOpen
  \bibfield  {author} {\bibinfo {author} {\bibfnamefont {E.}~\bibnamefont
  {Epelbaum}}, \bibinfo {author} {\bibfnamefont {H.}~\bibnamefont {Krebs}}, \
  and\ \bibinfo {author} {\bibfnamefont {U.~G.}\ \bibnamefont {Meißner}},\
  }\href {\doibase 10.1140/epja/i2015-15053-8} {\bibfield  {journal} {\bibinfo
  {journal} {Eur. Phys. J.}\ }\textbf {\bibinfo {volume} {A51}},\ \bibinfo
  {pages} {53} (\bibinfo {year} {2015}{\natexlab{b}})},\ \Eprint
  {http://arxiv.org/abs/1412.0142} {arXiv:1412.0142 [nucl-th]} \BibitemShut
  {NoStop}%
\bibitem [{\citenamefont {Binder}\ \emph {et~al.}(2016)\citenamefont {Binder}
  \emph {et~al.}}]{Binder:2015mbz}%
  \BibitemOpen
  \bibfield  {author} {\bibinfo {author} {\bibfnamefont {S.}~\bibnamefont
  {Binder}} \emph {et~al.} (\bibinfo {collaboration} {LENPIC}),\ }\href
  {\doibase 10.1103/PhysRevC.93.044002} {\bibfield  {journal} {\bibinfo
  {journal} {Phys. Rev.}\ }\textbf {\bibinfo {volume} {C93}},\ \bibinfo {pages}
  {044002} (\bibinfo {year} {2016})},\ \Eprint
  {http://arxiv.org/abs/1505.07218} {arXiv:1505.07218 [nucl-th]} \BibitemShut
  {NoStop}%
\bibitem [{\citenamefont {Hernández}\ \emph {et~al.}(2013)\citenamefont
  {Hernández}, \citenamefont {Nieves},\ and\ \citenamefont
  {Vicente~Vacas}}]{Hernandez:2013jka}%
  \BibitemOpen
  \bibfield  {author} {\bibinfo {author} {\bibfnamefont {E.}~\bibnamefont
  {Hernández}}, \bibinfo {author} {\bibfnamefont {J.}~\bibnamefont {Nieves}},
  \ and\ \bibinfo {author} {\bibfnamefont {M.~J.}\ \bibnamefont
  {Vicente~Vacas}},\ }\href {\doibase 10.1103/PhysRevD.87.113009} {\bibfield
  {journal} {\bibinfo  {journal} {Phys. Rev.}\ }\textbf {\bibinfo {volume}
  {D87}},\ \bibinfo {pages} {113009} (\bibinfo {year} {2013})},\ \Eprint
  {http://arxiv.org/abs/1304.1320} {arXiv:1304.1320 [hep-ph]} \BibitemShut
  {NoStop}%
\bibitem [{\citenamefont {Alvarez-Ruso}\ \emph {et~al.}(2016)\citenamefont
  {Alvarez-Ruso}, \citenamefont {Hernández}, \citenamefont {Nieves},\ and\
  \citenamefont {Vicente~Vacas}}]{Alvarez-Ruso:2015eva}%
  \BibitemOpen
  \bibfield  {author} {\bibinfo {author} {\bibfnamefont {L.}~\bibnamefont
  {Alvarez-Ruso}}, \bibinfo {author} {\bibfnamefont {E.}~\bibnamefont
  {Hernández}}, \bibinfo {author} {\bibfnamefont {J.}~\bibnamefont {Nieves}},
  \ and\ \bibinfo {author} {\bibfnamefont {M.~J.}\ \bibnamefont
  {Vicente~Vacas}},\ }\href {\doibase 10.1103/PhysRevD.93.014016} {\bibfield
  {journal} {\bibinfo  {journal} {Phys. Rev.}\ }\textbf {\bibinfo {volume}
  {D93}},\ \bibinfo {pages} {014016} (\bibinfo {year} {2016})},\ \Eprint
  {http://arxiv.org/abs/1510.06266} {arXiv:1510.06266 [hep-ph]} \BibitemShut
  {NoStop}%
\bibitem [{\citenamefont {Hernández}\ and\ \citenamefont
  {Nieves}(2017)}]{Hernandez:2016yfb}%
  \BibitemOpen
  \bibfield  {author} {\bibinfo {author} {\bibfnamefont {E.}~\bibnamefont
  {Hernández}}\ and\ \bibinfo {author} {\bibfnamefont {J.}~\bibnamefont
  {Nieves}},\ }\href {\doibase 10.1103/PhysRevD.95.053007} {\bibfield
  {journal} {\bibinfo  {journal} {Phys. Rev.}\ }\textbf {\bibinfo {volume}
  {D95}},\ \bibinfo {pages} {053007} (\bibinfo {year} {2017})},\ \Eprint
  {http://arxiv.org/abs/1612.02343} {arXiv:1612.02343 [hep-ph]} \BibitemShut
  {NoStop}%
\bibitem [{\citenamefont {Sobczyk}\ \emph {et~al.}(2018)\citenamefont
  {Sobczyk}, \citenamefont {Hernández}, \citenamefont {Nakamura},
  \citenamefont {Nieves},\ and\ \citenamefont {Sato}}]{Sobczyk:2018ghy}%
  \BibitemOpen
  \bibfield  {author} {\bibinfo {author} {\bibfnamefont {J.}~\bibnamefont
  {Sobczyk}}, \bibinfo {author} {\bibfnamefont {E.}~\bibnamefont {Hernández}},
  \bibinfo {author} {\bibfnamefont {S.}~\bibnamefont {Nakamura}}, \bibinfo
  {author} {\bibfnamefont {J.}~\bibnamefont {Nieves}}, \ and\ \bibinfo {author}
  {\bibfnamefont {T.}~\bibnamefont {Sato}},\ }\href {\doibase
  10.1103/PhysRevD.98.073001} {\bibfield  {journal} {\bibinfo  {journal} {Phys.
  Rev.}\ }\textbf {\bibinfo {volume} {D98}},\ \bibinfo {pages} {073001}
  (\bibinfo {year} {2018})},\ \Eprint {http://arxiv.org/abs/1807.11281}
  {arXiv:1807.11281 [hep-ph]} \BibitemShut {NoStop}%
\bibitem [{\citenamefont {Sato}\ \emph {et~al.}(2003)\citenamefont {Sato},
  \citenamefont {Uno},\ and\ \citenamefont {Lee}}]{Sato:2003rq}%
  \BibitemOpen
  \bibfield  {author} {\bibinfo {author} {\bibfnamefont {T.}~\bibnamefont
  {Sato}}, \bibinfo {author} {\bibfnamefont {D.}~\bibnamefont {Uno}}, \ and\
  \bibinfo {author} {\bibfnamefont {T.~S.~H.}\ \bibnamefont {Lee}},\ }\href
  {\doibase 10.1103/PhysRevC.67.065201} {\bibfield  {journal} {\bibinfo
  {journal} {Phys. Rev.}\ }\textbf {\bibinfo {volume} {C67}},\ \bibinfo {pages}
  {065201} (\bibinfo {year} {2003})},\ \Eprint
  {http://arxiv.org/abs/nucl-th/0303050} {arXiv:nucl-th/0303050 [nucl-th]}
  \BibitemShut {NoStop}%
\bibitem [{\citenamefont {Matsuyama}\ \emph {et~al.}(2007)\citenamefont
  {Matsuyama}, \citenamefont {Sato},\ and\ \citenamefont
  {Lee}}]{Matsuyama:2006rp}%
  \BibitemOpen
  \bibfield  {author} {\bibinfo {author} {\bibfnamefont {A.}~\bibnamefont
  {Matsuyama}}, \bibinfo {author} {\bibfnamefont {T.}~\bibnamefont {Sato}}, \
  and\ \bibinfo {author} {\bibfnamefont {T.~S.~H.}\ \bibnamefont {Lee}},\
  }\href {\doibase 10.1016/j.physrep.2006.12.003} {\bibfield  {journal}
  {\bibinfo  {journal} {Phys. Rept.}\ }\textbf {\bibinfo {volume} {439}},\
  \bibinfo {pages} {193} (\bibinfo {year} {2007})},\ \Eprint
  {http://arxiv.org/abs/nucl-th/0608051} {arXiv:nucl-th/0608051 [nucl-th]}
  \BibitemShut {NoStop}%
\end{thebibliography}%

\end{document}